\documentclass[useAMS,usenatbib]{mn2e}

\usepackage{psfig}

\def\etal{et al.}

\def\teff{\ifmmode T_{\rm eff} \else $T_{\mathrm{eff}}$\fi}

\def\ltsima{$\buildrel<\over\sim$}
\def\lsim{\lower.5ex\hbox{\ltsima}}

\newcommand{\hii}{H~{\sc ii}}
\newcommand{\ha}{\ifmmode {\rm H}\alpha \else H$\alpha$\fi}
\newcommand{\hb}{\ifmmode {\rm H}\beta \else H$\beta$\fi}
\newcommand{\lya}{\ifmmode {\rm Ly}\alpha \else Ly$\alpha$\fi}
\newcommand{\hei}{He~{\sc i}}

\def\micron{$\mu$m}

\def\ergs{erg s$^{-1}$}
\def\ergscm{erg s$^{-1}$ cm$^{-2}$}
\def\msun{\ifmmode M_{\odot} \else M$_{\odot}$\fi}
\def\msunyr{\ifmmode M_{\odot} {\rm yr}^{-1} \else M$_{\odot}$ yr$^{-1}$\fi}
\def\zsun{\ifmmode Z_{\odot} \else Z$_{\odot}$\fi}
\def\lsun{\ifmmode L_{\odot} \else L$_{\odot}$\fi}

\def\mup{\ifmmode M_{\rm up} \else M$_{\rm up}$\fi}
\def\mlow{\ifmmode M_{\rm low} \else M$_{\rm low}$\fi}
\def\vwfpc{V$_{\rm 606W}$}
\def\iwfpc{I$_{\rm 814W}$}
\def\zacs{z$_{\rm 850LP}$}
\def\jnic{H$_{\rm 110W}$}
\def\hnic{H$_{\rm 160W}$}

\def\aap{A\&A}

\def\aj{AJ}
\def\apj{ApJ}
\def\apjl{ApJL}
\def\apjs{ApJS}
\def\mnras{MNRAS}

\newcommand{\oh}{\ifmmode 12 + \log({\rm O/H}) \else$12 + \log({\rm
O/H})$\fi}
\def\Oiii{[O~{\sc iii}] $\lambda\lambda$4959,5007}
\def\hyperz{{\em Hyperz}}
\def\flyf{\ifmmode f_{\rm Lyf} \else $f_{\rm Lyf}$\fi}
\def\pz{\ifmmode P(z) \else $P(z)$\fi}
\def\ki2{\ifmmode \chi^2 \else $\chi^2$\fi}
\def\zphot{\ifmmode z_{\rm phot} \else $z_{\rm phot}$\fi}

\title[Stellar populations and \lya\ emission in two lensed $z\ga 6$ galaxies]
{Stellar populations and \lya\ emission in two lensed $z\ga 6$ galaxies} 

\author[Daniel Schaerer and Roser Pell\'o]{Daniel Schaerer$^{1,2}$
\thanks{E-mail: daniel.schaerer@obs.unige.ch} and
Roser Pell\'o$^2$\\
$^{1}$Observatoire de Gen\`eve,
51, Ch. des Maillettes, CH-1290 Sauverny, Switzerland\\
$^{2}$Laboratoire d'Astrophysique (UMR 5572),
Observatoire Midi-Pyr\'en\'ees,
14 Avenue E. Belin, F-31400 Toulouse, France}

\begin{document}
\date{}
%\date{Accepted 1988 December 15. Received 1988 December 14; 
%in original form 1988 October 11}
%\date{\large{\bf Draft - not for circulation!}}

%\pagerange{\pageref{firstpage}--\pageref{lastpage}} \pubyear{2002}

\maketitle

%\label{firstpage}

\begin{abstract}
We present an analysis of two strongly lensed galaxies at $z=6.56$
and $z \sim 7$
for which multi-band photometric and spectroscopic observations are available.
For one source the data include recent HST and Spitzer observations.
Using an SED fitting technique considering a number of parameters 
(various libraries of empirical and theoretical template spectra,
variable extinction and extinction laws)
we attempt to constrain the properties of their stellar populations
(age, star formation history, mass) and their intrinsic \lya\ emission.
The following main results are obtained for the individual galaxies:

\begin{itemize} 
\item {\bf Triple arc in Abell 2218, probable $z \sim 7$ galaxy:}
The most likely redshift of this source is $z \sim$ 6.0--7.2
taking into account both our photometric determination and lensing 
considerations.
SED fits indicate generally a low extinction ($E(B-V) \la 0.05$)
but do not strongly constrain the star formation (SF) history. 
Best fits have typical ages of $\sim$ 3 to 400 Myr. A reasonable
maximum age of (250--650) Myr (1 $\sigma$ interval) can be estimated . 
However, the apparent 4000 \AA\ break 
observed recently from combination of IRAC/Spitzer and HST observations,
can also equally well be reproduced with the template of a young
($\sim$ 3--5 Myr) burst where strong restframe optical emission lines 
enhance the 3.6 and 4.5 \micron\ fluxes.
The estimated SFR is typically $\sim$ 1 \msunyr\ for a Salpeter IMF
from 1-100 \msun, in agreement with previous estimates.
The unknowns on the age and star formation history could easily 
explain the apparent absence of \lya\ in this galaxy. 

\item {\bf Abell 370 HCM 6A:} 
The available \lya\ and continuum observations  indicate 
basically two possible solutions:
1) a young burst or ongoing constant SF with non-negligible extinction
or 2) a composite young + ``old''  stellar population.
In the first case one obtains a best fit $E(B-V) \sim 0.25$, or 
$A_V \sim$ 0.5--1.8 at a 1 $\sigma$ level.
In consequence we obtain SFR $\sim$ 11--41 \msunyr, higher than earlier
estimates, and we estimate a fairly high total luminosity ($L \sim
(1-4) \times 10^{11} \lsun$) for this galaxy, in the range of luminous infrared galaxies.
A \lya\ transmission of $\sim$ 23--90 \% is estimated from our best fit models.
Other properties (age, SF history) remain largely unconstrained.
In case of composite stellar populations the SFR, mass, and luminosity
estimate is lower.
The two scenarios may be distinguishable with IRAC/Spitzer observations at
3.6 and 4.5 \micron.

\end{itemize} 

\end{abstract}


%%%%%%%%%%%%%%%%%%%%%%%%%%%%%%%%%%%%%%%%%%%%%%%%%%%%%%%%%%%%%%%%
\section{Introduction}
\label{s_intro}

Little is known about the stellar properties, extinction, and
the expected intrinsic \lya\ emission of distant, high redshift 
galaxies. Indeed, although it has in the recent past become possible 
through various techniques to detect already sizeable numbers of 
galaxies at $z \ga 5$ 
(see e.g.\ the reviews of Taniguchi et al.\ 2003 and Spinrad 2004)
the information available on these objects remains generally scant.
For example, in many cases the galaxies are just detected in
two photometric bands and \lya\ line emission, when present, 
serves to determine the spectroscopic redshift (e.g.\ Bremer \etal\ 2004,
% Bouwens \etal\ 2004, --> has more than 2 bands !
Dickinson \etal\ 2004, Bunker \etal\ 2004).
Then the photometry is basically used to estimate the star formation rate 
(SFR) assuming standard conversion factors between the UV restframe
light and the SFR, and nothing is known about the extinction,
and the properties of the stellar population (such as age, detailed 
star formation histories etc.)

At higher redshift ($z \ga 6$) even less information is generally available
(but see a recent study of Eyles \etal\ 2005 on two $z \sim 6$ galaxies
observed with HST and Spitzer).
Many objects are found by \lya\ emission, but remain weak or sometimes even 
undetected in the continuum (e.g.\ Rhoads \& Malhotra 2001, Kodaira \etal\ 2003,
Cuby \etal\ 2003, Ajiki \etal\ 2003, Taniguchi et al.\ 2004).
In these cases the \lya\ luminosity can be determined 
and used to estimate a SFR using again standard conversion factors. 
Also the \lya\ equivalent width is estimated,
providing some possible clue on the nature of these sources.
However, this has lead to puzzling results e.g.\ for the 
sources from the LALA survey 
which seem to show unusually large \lya\ equivalent widths
that are difficult to understand without invoking exceptional
conditions (PopIII stars?; Malhotra \& Rhoads 2002, Rhoads \etal\ 2003). 
Given the few data available
for the LALA sources it is fair to say that the nature of these objects, 
their stellar populations, extinction etc.\ remain currently 
largely unknown (cf.\ Dawson \etal\ 2004).
When possible, a simple comparison between the UV and \lya\ SFR
is undertaken providing possibly information on the 
\lya\ transmission, i.e.\ the partial absorption of \lya\ photons
on their sight line through the intergalactic medium (e.g.\ Haiman
2002, Santos 2004) and/or on partial \lya\ ``destruction'' processes 
close to the source 
(e.g.\ due to dust or ISM geometry; Charlot \& Fall 1993, Valls-Gabaud 1993,
Tenorio-Tagle \etal\ 1999, Mas-Hesse \etal\ 2003).

Notable exceptions of $z \ga 4$ samples for which some estimate 
of extinction is available from multi-band photometry include 
work on the Subaru Deep Survey (Ouchi \etal\ 2004) 
and {\em GOODS} data (e.g.\ Papovich \etal\ 2004).
Lehnert \& Bremer (2004) also discuss some preliminary information
on little extinction in their $z>5$ sources.
Interestingly, in their study of a $z=5.34$ galaxy discovered by Dey \etal\ (1998), 
Armus \etal\ (1998) find indications for significant reddening
($A_V > 0.5$ mag) from analysis of the observed SED and from the presence
of \lya\ emission.

In a similar manner we will here present a consistent study of the
stellar population properties, extinction, and \lya\ emission for 
two galaxies at redshift $z \ga 6$.
For this aim we use two distant ($z \ga 6$) gravitationally lensed galaxies
for which multi-band photometry is available (detection in at least 3--4 bands).
Through a quantitative analysis of their SED, using a vast library 
of empirical and theoretical template spectra, we aim to constrain
properties of the stellar populations, such as age and star formation (hereafter SF)
history (burst or constant SF?) and their extinction.
Furthermore by comparing the  \lya\ emission expected from the
stellar population constraint with the observed \lya\ flux we
estimate consistently the \lya\ ``transmission'' for the individual sources.

The \lya\ transmission and SF properties derived here can in principle 
be used to infer the ionisation fraction of hydrogen in the IGM at a 
given redshift (cf.\ Haiman 2002, Santos 2004),
a key quantity of interest for the
study of the reionisation history of the Universe (cf.\ review from 
Barkana \& Loeb 2001).
Obviously the present ``exploratory'' work will have to be extended to
larger galaxy samples, and sophisticated tools will probably be
needed to interpret such results in terms of IGM properties 
(cf.\ Gnedin \& Prada 2004). However, this approach should be complementary to
other methods probing the reionisation history by
measuring the Gunn-Peterson optical depth observed in quasar spectra as 
a function of redshift (e.g.\ Becker \etal\ 2001, Fan \etal\ 2003), 
or by  comparing \lya\ luminosity functions at different redshifts 
(e.g.\ Malhotra \& Rhoads 2004).

The remainder of the paper is structured as follows.
In Sect.\ \ref{s_obs} we summarise the adopted observational constraints
from the literature.
Our modeling technique is described in Sect.\ \ref{s_models}.
The detailed results for each galaxy are presented in Sects.\ \ref{s_370}
and \ref{s_kesr}.
Our main conclusion are summarised in  Sect.\ \ref{s_conclude}.
%A brief discussion and a summary of our main conclusion are
%given in  Sects.\ \ref{s_discuss} and \ref{s_conclude}.

%%%%%%%%%%%%%%%%%%%%%%%%%%%%%%%%%%%%%%%%%%%%%%%%%%%%%%%%%%%%%%%%
\section{Observational constraints}
\label{s_obs}

The two galaxies studied here are:
1) The probable $z \sim 7$ galaxy recently discovered
by Kneib \etal\ (2004, hereafter KESR), which presently lacks of a spectroscopic
redshift but for which rather accurate multi-band HST observations are available,
allowing us in particular also to derive a fairly reliable photometric redshift.
2) the $z=6.56$ \lya\ emitter HCM 6A behind the lensing cluster Abell 370.
We now summarise the observational data, 
taken from the literature.
The adopted redshift and gravitational magnification factors
are listed in Table \ref{tab_props}.

Before proceeding let us mention for clarity that these two objects
are generally considered to be star forming galaxies (starbursts), not
AGN (narrow line - type II - or others), as no contradicting information
is available so far. However, one must bear in mind that some of the
interpretations presented below (and in the literature) may need to be revised,
should this assumption be incorrect.

{\bf Triple arc in Abell 2218:}
The observational data for this object, named Abell 2218 KESR hereafter, 
is taken from Kneib et al.\ (2004, hereafter KESR) and from Egami \etal\ (2005).
The photometry from KESR includes observations with HST (WFPC2, ACS, NICMOS) in
\vwfpc (undetected), \iwfpc, \zacs, and \hnic, and with NIRC/Keck
in $J$.
Subsequently, additional photometry was obtained with NICMOS/HST in the $J$ band
(F110W), and with IRAC/Spitzer at 3.6 and 4.5 \micron\ (see Egami et al.)
For our computations (see below) we have used the appropriate filter transmission
curves. In particular, 
updated transmission curves were used for the ACS and NICMOS filters
(M. Sirianni 2003, private communication; Sirianni et al.\ 2004; 
R.\ Thompson 2003, private communication).

Few brief comments concerning the photometry are needed here.
First, KESR present photometry for two multiple images
(a and b). 
Apparently sources a and b differ in the \zacs\ flux 
(with quoted errors of $\pm$ 0.05 mag) by 3.2 $\sigma$, 
whereas the fluxes in the other filters agree well within 1 $\sigma$. 
Differential lensing across the images together with sampling
  effects could be responsible for this small discrepancy. As
  we are interested in a global representative SED for this source, 
  we have chosen to use the averaged photometric SED, the magnification 
  factors being the same for the two images.
Finally, we have also noted some apparent discrepancies between
the measurements reported in KESR and Egami \etal, the most
important one being the \hnic\ flux, which is $\sim$ 15--20 \%
(3--4 $\sigma$) higher in the latter publication.
These differences are mostly due to the use of different apertures
on different repixeled/rescaled images (J.\ Richard, 2004, private
communication).
Again, this illustrates the difficulty in deriving reliable colors for
extended arcs.
To account for these small discrepancies and for the possible error 
underestimate we therefore adopt a minimum photometric 
error of 0.15 mag in all filters.
It is worth noting that photometric errors translate into absolute 
flux calibration errors for fitting purposes. 
As we will see below, adopting the latter minimum errorbars
significantly improves the SED fits.

In addition to the photometry, the non-detection of the source with Keck LRIS spectroscopy
provides an upper limit on the continuum flux between 9000 and 9300 \AA\ (KESR).
This upper limit will be used as an additional constraint in our SED modeling.
KESR also indicate a possible  drop of the continuum below $\sim$ 9800 \AA\ from
their Keck II NIRSPEC spectrum. For various reasons the reality of this
spectral break is questionable.
First a true neutral hydrogen break (``\lya\ break'') at $\sim$ 9800 \AA, far 
in the red wing of the \zacs\ filter, seems incompatible with the relatively strong
flux measured in this filter. Furthermore, test computations show that 
such a break is difficult if not impossible to reconcile with
our spectral modeling. In any case the significance of this finding
appears questionable as the detected continuum is extremely 
faint and noisy. The reality of this spectral feature is now also questioned 
by Egami \etal\ (2005). For these reasons this information is 
discarded from our spectral fitting.

In practice we have retained the following two variants to describe
the observed SED of this source:
{\em SED1)} The average fluxes (\iwfpc, \zacs, \jnic, \hnic) of images a and b 
from KESR plus the IRAC/Spitzer data of image b from Egami et al. 
{\em SED2)} All fluxes from image b from  Egami et al.
%{\em SED3)} Same as 2, except for the \hnic\ flux taken as the average of a and b
%from KESR.
These are treated as SEDs from two different objects.
Furthermore, for each of these ``objects'' we have computed two cases 
in our SED fitting described below: 
{\em i)} The observed SEDs in \iwfpc, \zacs, \jnic, \hnic, 3.6, and 4.5 \micron.
{\em ii)} Same as (i) plus the flux limits from the \vwfpc\ and
Keck LRIS non-detections.

No emission line has so far been detected for Abell 2218 KESR.
Its spectroscopic redshift remains therefore presently unknown
but the well-constrained mass model for the cluster strongly suggests 
a redshift $z \ga$ 6 for this source (KESR, Egami \etal\ 2005).
The magnification factors of both images a and b is $\mu=25 \pm 3$,
according to KESR.

{\bf Abell 370 HCM6A:}
The observational data of this $z=6.56$ galaxy is taken from Hu et al.\ (2002).
The photometry includes $VRIZJHK^\prime$ from Keck I and II
(LRIS and Echellette Spectrograph and Imager) and from Subaru
(CISCO/OHS).
%The redshift identification/confirmation, $z=6.56$, is based on an 
%asymmetric emission line identified as \lya.
The gravitational magnification of the source is $\mu=4.5$
according to Hu et al.\ (2002).

Photometric fluxes and errors were adopted from their Fig.\ 3.
Where possible the appropriate filter transmission curves 
were used.
%available on the Web were used\footnote{
%{\tt http://alamoana.keck.hawaii.edu/inst/lris/filter_list.html},
%{\tt http://alamoana.keck.hawaii.edu/inst/lris/ccd2_qe.html},
%{\tt http://www.naoj.org/Observing/Instruments/OHS/camera/filters.html}}.
%
The ``$Z$'' band filter transmission is somewhat uncertain, as
these observations were undertaken using an RG850 filter,
which together with the LRIS optics and the CCD response,
yields a transmission similar to a $Z$ band filter (Hu et al.\ 1999).
Our approximate filter curve shows a blueward shift of $\lambda_{\rm eff}$
by $\sim$ 200 \AA\ compared to the information given by Hu et al.\ (1999).
%The synthesized ``$Z$'' band flux may be somewhat affected by
%this uncertainty. 
However, since the redshift is known for this source and since
we adjust the observed flux (not magnitude) in this band, 
this should not affect our conclusions.

%%%%%%%%%%%%%%%%%%%%%%%%%%%%%%%%%%%%%%%%%%%%%%%%%%%%%%%%%%%%%%%%
\section{SED modeling}
\label{s_models}

% % % % % % % % % % % % % % % % 
\begin{figure}
\centerline{\psfig{figure=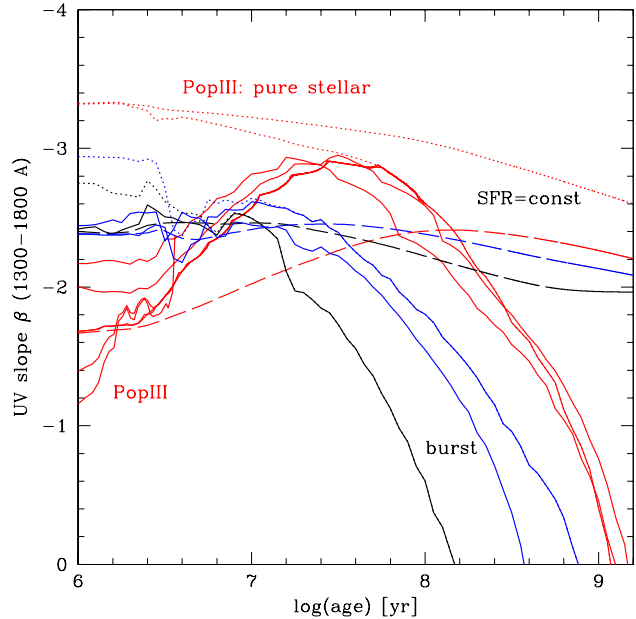,width=8.8cm}}
% from macro: hyperz
\caption{Temporal evolution of the UV slope $\beta$ measured between 1300 and 1800
\AA\ from synthesis models of different metallicities and for instantaneous
bursts (solid lines) and constant SF (long dashed lines).
Black lines show solar metallicity models, red lines metallicities between 
$Z = 10^{-5}$ and zero (PopIII), blue lines intermediate cases of $Z=0.004$ and 0.0004. 
The dotted lines show $\beta$ if nebular continuous emission is neglected,
i.e.\ assuming pure stellar emission.
Note especially the strong degeneracies of $\beta$ in age and metallicity for
bursts, the insensitivity of $\beta$ on $Z$ for constant SF, and
the rather red slope for young very metal-poor bursts. Further discussions in the text
}
\label{fig_beta}
\end{figure}
% % % % % % % % % % % % % % % % 

\subsection{Main restframe UV-optical SED features of high-z galaxies
and their ``information content''}
\label{s_uv}

Before proceeding to the fits of the individual SEDs a brief comment
on the available SED features seems appropriate.

For obvious reasons the available SED (basically from broad-band photometry)
of high-z ($z \ga 6$) galaxies is primarily limited to the rest-frame UV 
(when observed from the ground) or optical spectrum (when available e.g.
with Spitzer and future satellite missions).
The main information ``encoded'' in this SED is therefore:
{\em 1)} the neutral HI break shortward of \lya\ (hereafter the ``\lya'' break)
   due to the strong or complete Gunn-Peterson trough,
{\em 2)} the slope of the UV spectrum, and 
{\em 3)} possibly a 4000 \AA\ break (hereafter denoted Balmer break), 
if present and covered by the observations.
In addition the presence of the \lya\ line, mostly used to determine
spectroscopically the redshift, provides clear evidence for ongoing massive 
star formation (hereafter SF).

The position of the \lya\ break depends essentially on redshift.
The UV slope depends on the intrinsic spectrum -- in turn depending
mostly on age and SF history -- and on the extinction, i.e.\ the extinction
law and the amount of reddening.
The Balmer break becomes visible (in absorption) in the continuum of
stellar populations after $\ga$ 10--30 Myr.
\lya\ emission, if due to stellar photoionisation and not AGN activity,
indicates the presence of young ($\la$ 10 Myr) massive ionizing stars. 

Concerning the UV slope, it is useful to recall that this quantity
\footnote{Various definitions of the UV slope exist. The most commonly
used ones, generally denoted $\beta$, are defined as the power-law
index of the SED in $F_\lambda$ versus $\lambda$ over a certain
wavelength interval.} does not lend itself to determine
the metallicity of a star forming galaxy from a theoretical
point of view and in terms of individual objects. 
The reasons are that intrinsically the UV slope shows only small variations
with metallicity, and that the slope depends strongly on the exact
SF history (see e.g.\ Leitherer \& Heckman 1995, Meurer \etal\ 1995).
This is illustrated in Fig.\ \ref{fig_beta} where $\beta$ (measured over
the interval 1300-1800 \AA) is plotted as a function of age for
populations of metallicities between solar and zero (PopIII) and for
the limiting cases of bursts and SFR=const.
Furthermore, as pointed out in Schaerer (2002, 2003) and also shown in this
figure, for very low metallicities ($Z\la 1/50\zsun$) nebular continuous 
emission becomes dominant even 
down to UV wavelengths (longward of \lya), such that the observed
integrated (stellar+nebular) spectrum has even a flatter UV slope
than high metallicity starbursts.
In other words, even for bursts, there are strong intrinsic degeneracies 
of $\beta$ between age and metallicity, to which the additional effect
of reddening must be added,
including the uncertainties on the a priori unknown extinction law.
It is therefore evident that on an individual object basis 
there are in general degeneracies between age, metallicity, SF history,
and extinction.
However, this does not preclude the possible existence of 
statistical correlations between quantities such as e.g.\ $\beta$
and metallicity in large samples of galaxies, as known to
hold e.g.\ for local UV selected starbursts (cf.\ Heckman et al.\ 1998).
Also, as we will see below, there are cases where the UV slope
and the mere fact of the presence of an emission line allow us nevertheless to
lift some degeneracies and therefore to determine interesting
constraints on the stellar population and on extinction.

The behaviour of the 4000 \AA\ break and its use as an age indicator has
extensively been discussed in the literature (e.g.\ Bruzual 1983, and
recently Kauffmann \etal\ 2003).
As in simple stellar populations its amplitude is basically a monotonically 
increasing function of age, an estimate of the break,  e.g.\ obtained from 
3.8-4.5 \micron\ photometry with IRAC/Spitzer and JHK photometry in $z\ga 6$
galaxies, provides information on the age of the light emitting stellar population.
Since the exact SF history cannot be determined in these cases
(in contrast to studies at low $z$, cf.\ Kauffmann \etal) the amplitude
of the break provides a range of ages, the minimum age being given by
instantaneous bursts, the maximum age from models with constant SF.
For obvious reasons this maximum ``luminosity weighted'' age derived from a 
measure of the Balmer break is also unaffected by considerations of 
possible multiple stellar populations. The same is not true for the minimum age,
which is, however, of less cosmological interest.

\subsection{Spectral fitting}
\label{s_fit}
For the spectral fitting we use a slightly adapted version 
of the photometric redshift code \hyperz\ 
%developed by R.\ Pell\'o and collaborators 
of Bolzonella et al.\ (2000).
\hyperz\ does standard SED fitting using a number of modeling parameters.
The free parameters for the SED modeling are:

\begin{itemize}
\item [{\em 1)}] the spectral template, 
\item [{\em 2)}] extinction and the reddening law, 
\item [{\em 3)}] a parameter \flyf\ describing possible deviations from the
average Lyman forest attenuation from Madau (1995).
\end{itemize}
For Abell 2218 KESR the source redshift is also a free parameter.

For the spectral templates we use a large compilation of empirical
and theoretical SED, including starbursts, QSO, and galaxies of all Hubble types, 
and covering various star formation histories (bursts, exponentially decreasing, 
constant SF) and various metallicities.
For most applications we group the templates in the following way:
\begin{itemize}
\item {\bf Starbursts and QSOs (hereafter SB+QSO):} this group includes the starburst 
templates with $E(B-V)$ from $<0.1$ to 0.7 from the Calzetti et al.\ (1994) and 
Kinney et al.\ (1996) atlas, 
the HST QSO template of Zheng et al.\ (1997), as well as UV-optical spectrum 
of the metal-poor galaxy SBS 0335-052 with numerous strong optical emission lines
(and an extinction of $E(B-V) \sim 0.09$, Izotov \& Thuan 1998)
kindly communicated to us by Yuri Izotov (2002, private communication)

\item {\bf BCCWW+:} Bruzual \& Charlot (1998, private communication; 
cf.\ Bruzual \& Charlot 1993) evolving synthesis models
assuming bursts, constant star formation, and exponentially decaying star
formation histories 
reproducing present day spectra of galaxies of various types 
(E, S0, Sa, Sb, Sc, Sd, and Im) plus the empirical 
E, Sbc, Scd, and Im templates from Coleman et al.\ (1980),
as included in the public \hyperz\ version.
\item {\bf S03+:} Theoretical templates of starburst galaxies 
from Schaerer (2003) covering metallicities of $Z=0.02$ (solar), 0.008,
0.004, 0.001, 1/50 \zsun, $Z=10^{-5}$, $10^{-7}$, and zero metallicity
(PopIII). For low metallicities ($Z \le 10^{-5}$) these templates have 
been computed for 3 different assumptions on the IMF. The spectral
library includes burst models and models with a constant star formation rate
(SFR). For more details see Schaerer (2003). 
For the present work these computations were extended to cover
ages of up to 1 Gyr. These SEDs are available on request from the
first author and on the Web\footnote{{\tt http://obswww.unige.ch/sfr}}.
\end{itemize}

The standard extinction law adopted here is the one from 
Calzetti et al.\ (2000) determined empirically from nearby starbursts.
We also explore the possible implications of other laws, such as
the Galactic law of Seaton (1979) including the 2200 \AA\ bump,
and the SMC law from Pr\'evot \etal\ (1984) and Bouchet \etal\ (1985)
showing no UV bump, but a steeper increase of the extinction 
in the UV compared to Calzetti \etal. 

For the Lyman forest attenuation, \hyperz\ follows Madau (1995).
However, we allow for possible deviations from the mean attenuation
by varying the Lyman forest optical depths $\tau_{\rm eff}^{\alpha,\beta}$
by a multiplicative factor taking the values of (\flyf, 1., and $1/\flyf$).
Typically we adopted $\flyf=$ 2 or 3. Here $\tau_{\rm eff}^{\alpha,\beta}$
stands for the optical depths corresponding to the absorption between
\lya\ and Ly$\beta$, and between Ly$\beta$ and the Lyman limit respectively.

The following other minor changes have been made in our version (1.3ds) of
\hyperz. The calculation of the synthetic photometry deals correctly
with templates including strong spectral lines (emission or absorption).
Furthermore we make sure to use the proper filter transmission curves
usually given in photon units. Earlier versions of \hyperz\ and 
other codes (e.g.\ evolutionary synthesis codes) assume sometimes
(for ``historical'' reasons) that transmission curves be given in 
flux units. In case of wide filters, e.g.\ such as some ACS/HST 
filters, this may lead to small differences.
Other modifications concern essentially features related to the
user interface (additional outputs etc.).

 For given choices of the above parameters, the\hyperz\ code
   performs a standard minimisation fit to the observed SEDs and
   determines, for each point in the parameter space, the
   corresponding $\chi^2$ value. Using these $\chi^2$ values, it is
   possible to quantify the probabilities for the main free
   parameters, namely extinction, age of the spectral template, SF
   history, etc. When the SED fitting is based on theoretical
   templates, the SFR value is easily obtained and allows us to
   compare the expected values for the \lya\ flux to the actual ones.

To convert the observed/adjusted quantities to absolute values
we adopt the following cosmological parameters:
$\Omega_m=0.3$, $\Omega_\Lambda=0.7$, and 
$H_0 = 70$ km s$^{-1}$ Mpc$^{-1}$.

%%%%%%%%%%%%%%%%%%%%%%%%%%%%%%%%%%%%%%%%%%%%%%%%%%%%%%%%%%%%%%%%%
\section{Results for Abell 2218 KESR}
\label{s_kesr}

% % % % % % % % % % % % % % % % % % % % % % % % % % % % % % % % %
\begin{figure}
\centerline{\psfig{figure=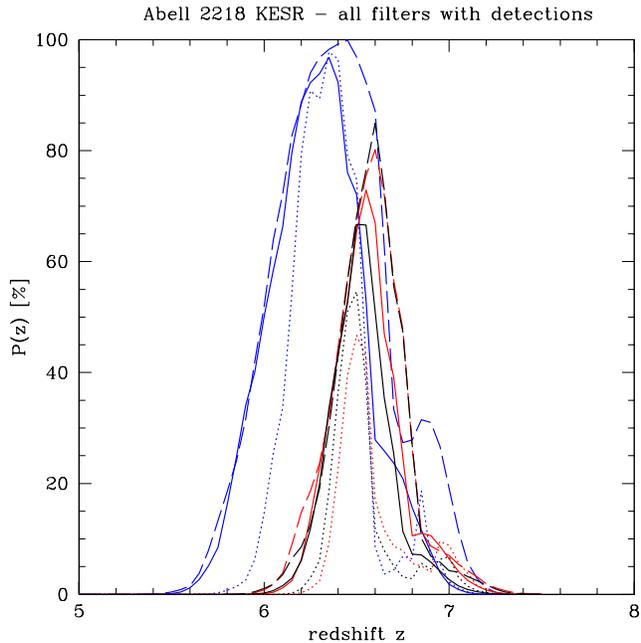,width=8.8cm}}
\caption{Photometric redshift probability
distributions \pz\ of Abell 2218 KESR using
three spectral template groups. Solid line: BCCWW+ template group, 
dotted: SB+QSO, long dashed: S03+. 
The three upper blue curves stand for the average SED of a and b (SED1), the lower red ones 
for the SED of object b (SED2) from Egami et al.\ (2005). 
%in red and SED3 in black.
In all cases a minimum photometric error of 0.15 mag was adopted.
The \pz\ shown here has been computed  based on all filters in which the object 
is detected (\iwfpc\ to 4.5 \micron).	}
\label{fig_pz_7}
\end{figure}
% % % % % % % % % % % % % % % % % % % % % % % % % % % % % % % % %

\subsection{Photometric redshift estimate}

As a spectroscopic redshift has not been obtained (yet) for this galaxy
we here examine its photometric redshift estimate.
In Fig.\ \ref{fig_pz_7} we show the photometric redshift probability
distributions \pz\ for the two SEDs (SED1, SED2) of Abell 2218 KESR described above
using the three spectral template groups and adopting a minimum photometric error
of 0.15 mag.
For each redshift, \pz\ quantifies the quality of the best fit model
obtained varying all other parameters (i.e.\ extinction, \flyf, spectral template among
template group).
Given the excellent HST (WFPC2, ACS and NICMOS) photometry, \pz\
is quite well defined: the photometric redshift ranges typically between 
$z_{\rm phot} \sim$ 5.5 and 7.3. Outside of the plotted redshift range
\pz\ is essentially zero.
If we assume the (smaller) quoted formal photometric errors (but note the discrepancies
discussed in Sect.\ \ref{s_obs}) \pz\ becomes more peaked, i.e.\ the photometric
redshift better defined. This is driven by the error on the \zacs\ flux, which determines 
the red side of the ``\lya'' break.  However, the resulting best fit value \zphot\ does not
change much. Furthermore, the fit quality is considerably decreased. 
This demonstrates the interest of such high accuracy measurements and the need for
reliable error estimates.

The predicted redshift distribution is found to be quite insensitive
to the exact template (as shown in Fig.\ \ref{fig_pz_7}), 
to the exact value of \flyf, and to the adopted extinction law 
(variations of the latter two are not shown).
However, we note that for this object the fits (and \pz) are 
improved when allowing for deviations from the average Madau (1995)
attenuation law. The curves shown here have been computed for $\flyf=2$.

More important in determining \pz\ is the exact SED.
As seen from Fig.\ \ref{fig_pz_7} the use of SED1 or SED2 lead to somewhat different 
\pz\ distributions. SED2 (cf.\ Sect.\ \ref{s_obs}) yields a somewhat larger redshifts, 
albeit with a somewhat yreduced fit quality.
These differences illustrate how uncertainties and difficulties in the photometric
measurements of such faint sources, whose origin are briefly discussed in 
Sect.\ \ref{s_obs}, propagate to the photometric redshift estimate.

All our best-fit solutions have redshift $\zphot \sim$ 6.25--6.63, 
lower than the redshift range estimated by KESR, 
but compatible with the more recent quantitative analysis of Egami et al.\ (2005).
%
%Understanding the origin of this difference is a useful exercise.
%It mostly stems from the fact 
%%This difference is mostly due to the fact
%that we compute the synthetic photometry taking the detailed transmission
%curves into account allowing us to quantify the SED and in particular
%its stringent spectral break between \iwfpc\ and \zacs. 
%In contrast, KESR estimated e.g.\ $6.6 < z < 7.1$ from a simple 
%estimate of the position of the 1216 \AA\ (\lya) break at the red end
%of the \iwfpc\ filter. In fact, although this filter basically ``ends''
%at $\sim$ 9850 \AA\ (corresponding to $z(\lya) \sim 7.1$), its effective
%wavelength is considerably smaller ($\lambda_{\rm eff} \sim$ 8066 \AA), its
%transmission is tilted towards $\lambda <\lambda_{\rm eff}$, and its 
%half-surface is $\sim$ 800 \AA (cf.\ Fig.\ 3 of KESR). 
%In other words, most of the ``weight'' of the \iwfpc\ filter is at 
%wavelengths $\la$ 8866 \AA.
%
%%ds
%As already mentioned, the subsequent analysis of Egami et al.\ 
%(2005) dropping these simplifications yields a somewhat lower 
%photometric redshift than KEST, in fair agreement with our results.
%%For these reasons, our quantitative photometric redshift
%%estimates lead to a smaller but still compatible redshift than 
%%found by KESR. 
%%Our photometric redshift estimates are quite compatible with
%%the quantitative analysis of Egami et al.\ (2005).
%
Given the various free parameters, uncertainties on the intrinsic SED, etc.
we conclude that the redshift of Abell 2218 KESR
is likely $z \sim$ 6.0 -- 7.2.
% but possibly also somewhat lower ($z \ga 5.6$) 
taking into account both our photometric determination 
and the lensing considerations of KESR.

\subsection{SED fits and inferences on the stellar population and on \lya.}
A large number of models have been computed using the different variants
of the SEDs describing this object (SED1-2), the different
filter combinations (non-detections + spectroscopic constraint) 
discussed in Sect.\ \ref{s_obs}, and varying the various model
parameters.
We first discuss briefly the main salient results
with the help of some illustrations. A more general discussion
of the results and their dependence on various assumptions follows.

\subsubsection{Age, star formation history, and extinction}

% % % % % % % % % % % % % % % % % % % % % % % 
\begin{figure}
%[hbt]
\centerline{\psfig{figure=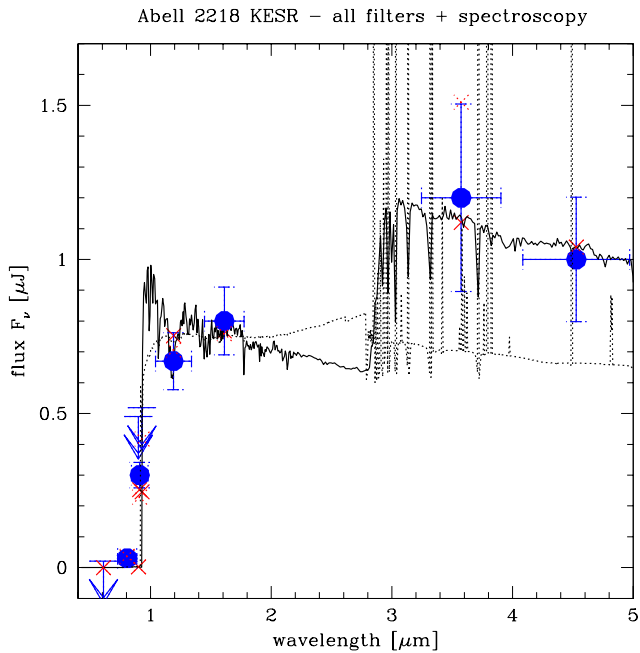,width=8.8cm}}
\caption{Best fits SEDs to the observations of Abell 2218
(SED2 from Egami \etal\ 2005, including the flux limit from the
non-detections in \vwfpc\ and at 9000-9300 \AA\ from spectroscopy;
cf.\ Sect.\ \ref{s_obs}).
The red crosses indicate the corresponding model broad band fluxes.
The solid line shows the best fit for a template from the S03+ group,
and dotted from the SB+QSO group.
The redshift for these solutions are $z \sim$ 6.63 and 6.54 respectively.
See text for more information}
\label{fig_7rev_all}
\end{figure}

\begin{figure}
%[hbt]
\centerline{\psfig{figure=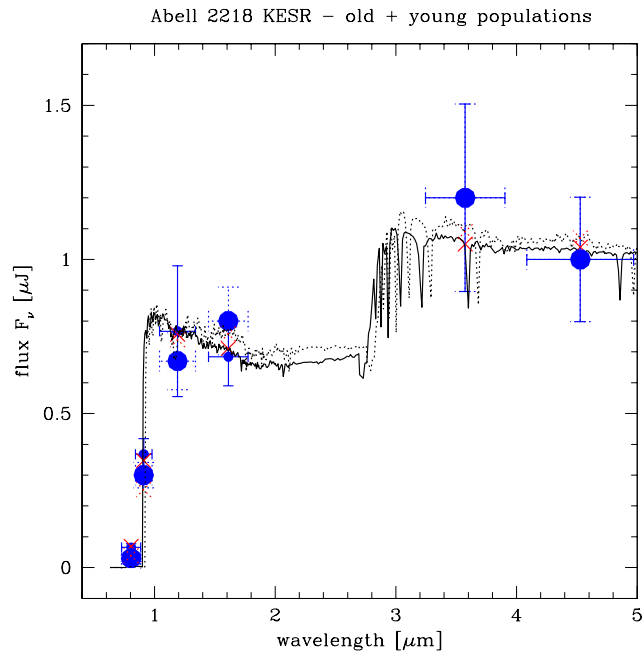,width=8.8cm}}
\caption{Best fits SEDs to the observations of Abell 2218
(Large symbols: SED2 from Egami \etal\ 2005; small symbols: SED1;
cf.\ Sect.\ \ref{s_obs}). Only true detections are taken into account.
The red crosses indicate the corresponding model broad band fluxes.
The solid (dotted) line shows the best fit for a template with constant
SFR from the S03+ group to the SED1 (SED2).
%The best fit redshifts are $\zphot = 6.40$ and 6.57 respectively.
The best ``maximum age'' fits correspond
to ages of 500 and 400 Myr, no extinction, and redshifts \zphot\ $\sim$ 6.40
and 6.57 respectively for SED1 and SED2.
See text for more information}
\label{fig_7rev_old}
\end{figure}
% % % % % % % % % % % % % % % % % % % % % % % 

Figure \ref{fig_7rev_all} shows the best fit models
to the SED2 including the upper limits from the \iwfpc\ and
LRIS spectroscopy for the S03+ and SB+QSO template groups.
The best fit redshifts are \zphot\ $=6.63$ and 6.54 respectively.
These fits show in particular that the spectroscopic constraint can be
accommodated simultaneously with the observed \zacs\ flux;
the resulting fits are within the 1 $\sigma$ errors in all bands.
The best fit from the S03+ group corresponds to a burst with
an age of 15 Myr at solar metallicity and no extinction (solid line).
Similarly good fits are also obtained for lower metallicity.
The best fit with empirical starburst and QSO templates (SB+QSO) 
is obtained with the spectrum of the metal-poor \hii\ galaxy SBS 0335-052
(dotted line in Fig.\ \ref{fig_7rev_all}).
In this case the apparent Balmer break observed between the NICMOS/HST and 
IRAC/Spitzer domain is simply explained by the presence of strong
emission lines in the 3.6 and 4.5  \micron\ filters
\footnote{The main lines are between H$\gamma$, \hb\ and \Oiii\
in the 3.6 \micron\ filter and \hei\ $\lambda$5876 in the 4.5 \micron\ filter.
E.g.\ for the emission lines between H$\gamma$ and \Oiii\ 
the total observed equivalent width (boosted by the $(1+z)$ factor)
is $\sim$ 9130 \AA, as estimated from the data of Izotov \& Thuan (1998),
compared to a filter surface of $\sim$ 6600 \AA.}
and some additional extinction to reduce the restframe UV flux.
The extinction needed is $A_V = 0.6$ for the Calzetti \etal\ law, or
$A_V = 0.2$ for the Pr\'evot \etal\ extinction law.
In terms of age the restframe UV to optical spectrum (continuum and lines) of 
SBS 0335-052 corresponds to a young population of $\sim$ 3--5 Myr according to 
the analysis of Papaderos \etal\ (1998) and Vanzi \etal\ (2000). 
Of course, the presence of an older population in addition to the starburst cannot be 
excluded on the present grounds.
In short, the observed SED of Abell 2218 KESR can be explained by a young population 
withour or with emission lines.
Spectroscopy in the 3--4 \micron\ range would be needed to distinguish the latter solution
from others.

%SBS all lines from 4300-5007: I/Hb~6. ==> with W(Hb)=217 ==> W(all)~1200
% W_obs(all)=W(all)*(1+z) ~ 8880 Ang !
% IRAC band 3.6: surface=6578
% IRAC band 4.5: surface=8850

Alternatively, good SED fits are also obtained with relatively ``old'' populations.
The oldest ages are obtained when invoking the longest SF timescale, i.e.\
constant SF. In this case the UV restframe flux remains high (due to the continuous
formation of massive stars) and older ages need to be attained to build up
a sufficient population of evolved stars with strong Balmer breaks, 
in order to reproduce the observed break.
This case is shown in Fig.\ \ref{fig_7rev_old} with fits with the S03+ templates
to the observed SED1 and SED2. The best fits solutions obtained here correspond
to ages of 500 and 400 Myr, no extinction, and redshifts \zphot\ $\sim$ 6.40
and 6.57 respectively. Similar, but somewhat older ages are obtained for
metallicities $Z < 0.008$, below the ones shown here.

% % % % % % % % % % % % % % % % % % % % % % % 
\begin{figure}
%[hbt]
%\centerline{\psfig{figure=contour_7rev.z655final.eps,width=8.8cm}}
\centerline{\psfig{figure=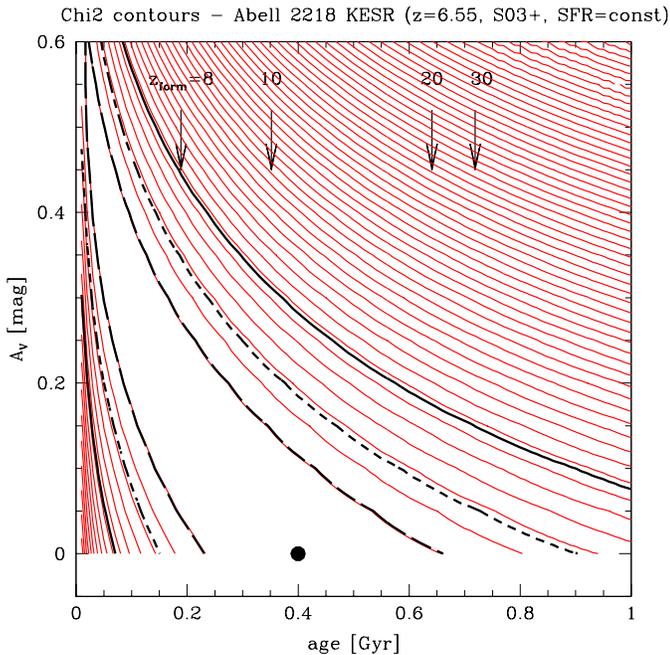,width=8.8cm}}
\caption{$\chi^2$ contour plot in extinction -- age for solutions 
fitting the observed SED2 at redshift $z=6.55$ with templates from the S03+ group
assuming constant SF.
The best fit solution is indicated by the black dot.
Equidistant $\chi^2$ levels with a spacing of 1 are shown. 
The (1D) 68, 90, and 99 \% confidence regions ($\Delta \chi^2 = 1.$, 2.71, 6.63) 
are delimited by the thick black lines in long dashed, dotted and solid
respectively.
Corresponding formation redshifts $z_{\rm form}$ (assuming the cosmological
parameters given in Sect.\ \ref{s_fit}) are indicated by the arrows.
Discussion in text}
%	# 	1D-68% Delta(chi2)=1.   corresponds to 1/5 here
%	# 	1D-90% Delta(chi2)=2.71 corresponds to 
%	# 	1D-99% Delta(chi2)=6.63 corresponds to \label{fig_7rev_contour}
\label{fig_7rev_contour}
\end{figure}
% % % % % % % % % % % % % % % % % % % % % % % 

A quantitative examination of the  ``maximum age'' allowed by the observations
(here SED2) is presented in Fig.\ \ref{fig_7rev_contour}, showing $\chi^2$ contours in the 
extinction--age plane for a given set of spectral templates (S03+ group with 
constant SFR), the Calzetti \etal\ extinction law, and a fixed redshift of
$z=6.55$. For these conditions the best fit is corresponds to 
400 Myr, zero extinction, and $\zphot=6.57$ (see Fig.\ \ref{fig_7rev_old}).
This Figure shows that a maximum age of $\sim$ (250--650) Myr (1 $\sigma$ interval)
is obtained, in good agreement with the modeling of Egami \etal\ (2005).
If true, this would correspond to a formation redshift of $z_{\rm form}
\sim$ 8.7--20 for our adopted cosmological parameters.
As clear from Fig.\ \ref{fig_7rev_contour}, even with constant SF models, 
younger populations with some extinction can also fit, although less well, 
the present observations.
However, solutions with low or zero extinction are generally preferred.

When varying the star formation history between these extreme cases (burst or SFR$=$const),
i.e.\ considering e.g.\ exponentially declining SF histories, any intermediate 
age can be found for obvious reasons. 
Such cases are e.g.\ obtained when fitting templates 
from the Bruzual \& Charlot models (not shown here) 
with exponentially decreasing SF histories and can be found in Egami et al.
As discussed in Sect.\ \ref{s_uv}, considering multiple stellar populations
(cf.\ Eyles \etal\ 2005) does not alter the above estimate of the maximum age
determined from constant SFR models.
In any case, the data available here does not allow us to constrain the SF history
and age further.

\subsubsection{\lya\ emission}
The observations obtained so far have not revealed any emission line
from this object (KESR). In particular \lya\ emission is lacking, which
could be puzzling for a source with intense star formation. 
As we have just seen a variety of star formation histories and ages
are possible for Abell 2218 KESR. Therefore one may or may not expect 
intrinsic \lya\ emission. 

A simple explanation for the apparent absence
of \lya\ emission could be to invoke an advanced age (in the post starburst phase).
However, even with a young age it is not necessary that \lya\ emission
be observed. E.g.\ the spectrum of the metal-poor \hii\ galaxy SBS 03335-052
which provides an excellent fit to the observed SED and shows strong emission
lines (cf.\ Fig.\ \ref{fig_7rev_all}) does not show \lya\ emission 
(Thuan \etal\ 1997, Kunth \etal\ 2003).
Alternatively, if intrinsically present, the \lya\ non-detection could 
be due a variety of factors: 
a redshift $z \la 6.4$ placing \lya\ below the spectral
range discussed in detail by KESR\footnote{This is probably excluded as, 
according to J.-P.\ Kneib (2005, private communication),
no emission line was found in the blue part of the spectrum taken with LRIS.},
a flux below their strongly
varying detection threshold, or other factors depressing the
\lya\ emission within the host galaxy (dust, ISM+HI geometry) and
in the intervening IGM.

In conclusion, from the available data the apparent lack of \lya\ emission from 
this source is not puzzling. However, it is not completely excluded
that the galaxy truely shows \lya\ emission, which has so far eluded detection.

\subsubsection{General comments on fits and discussion}
After these main findings we shall now quickly mention more ``technical''
results about the influence of various fit parameters.

Quite generally, the results on the age, SF history, extinction etc.\ depend little
on the different variants of the observed SEDs (SED1-2), on the inclusion or not of 
the non-detections in the fits, and on the use of the published formal errors or a 
minimum error of 0.15 mag (cf.\ Sect.\ \ref{s_obs}).
The results discussed above are therefore quite robust with respect to these
assumptions.
Small differences in the best fit values can, however, be obtained. E.g.\ the best fit
photometric redshift can vary by up to $\la$ 0.2 depending on adopting
SED1 or SED2.
%SED1-3.
In all cases we note that SED1 allows better fits (smaller $\chi^2$) than SED2.
% or SED3.
Adopting $\sigma_{\rm min}=0.15$ mag also significantly increases the fit quality.
Finally, considering variations around the mean Lyman forest attenuation improves the fits 
(especially as the HST photometry determining the ``\lya\ break'' is quite accurate).
In practice all best fits are found with an increased Lyman-forest opacity ($\flyf=2$).

To summarise, given the absence of a spectroscopic redshift, a fair number of 
good fits is found to the observations of Abell 2218 KESR
when considering all the free parameters.
The main conclusions from these ``best fits'' are:
\begin{itemize}
\item[{\em 1)}] Generally the determined extinction is negligible or zero
	quite independently of the adopted extinction law.
	The best fit with the empirical starburst spectrum of SBS 0335-052
	represents an exception to this case, requiring an additional
	$A_V \sim$ 0.2--0.6 mag, depending on the adopted extinction law.
\item[{\em 2)}] Although generally burst models fit somewhat better
	than those with constant star formation among the theoretical
	templates (BC, S03+), the data does not strongly constrain
	the star formation history.
\item[{\em 3)}] Typical ages between $\sim$ 15 and 400 Myr are obtained.
	A reasonable 1-$\sigma$ upper bound on the age of $\sim$ 650 Myr can be 
	obtained assuming constant star formation.
	However, the data can also be well fit with 
	%a template of 
	a very young ($\sim$ 
	3--5 Myr) stellar population with strong emission lines (using e.g.\ 
	the spectrum of the metal-poor galaxy SBS 0335-052). In this 
	case the apparent Balmer break observed between the HST and Spitzer 
	broad-band photometry is simply due to the presence of strong emission 
	lines affecting the red 3.6 and 4.5 \micron\ filters.
\item[{\em 4)}] Given degeneracies of the restframe UV spectra between age
  	and metallicity (cf.\ above) no clear indication on the galaxian
	metallicity can be derived, in contrast to the claim of KESR.
	Good fits to the available data can even be 
	found with solar metallicity starburst templates.
\item[{\em 5)}] Depending on the star formation history and age
	one may or may not expect intrinsic \lya\ emission, i.e.\
	an important \hii\ region around the object.
	The apparent absence of observed \lya\ emission does therefore
	not provide much insight.
\end{itemize} 

A more complete error analysis beyond the level presented here 
is difficult to achieve for a variety of reasons and clearly beyond the 
scope of this publication.

\subsubsection{SFR, stellar mass and luminosity}

The theoretical templates can also be used to estimate the 
stellar mass involved in the starburst or the star formation
rate when constant star formation is assumed. For this aim we 
use all the best fits to the three SEDs (SED1-3)
with the S03+ templates, we assume a typical redshift of 
$z=6.6$, and the magnification $\mu=25$ determined by KESR.
For the adopted cosmology the distance luminosity is then
$d_L=$64457.8 Mpc.
% ==> 4*pi*d_L^2 = 4.79e+59 cm^2

When constant SF is assumed one obtains the following star formation
rate: $SFR \sim (0.9-1.1)$ \msunyr\
(for a Salpeter IMF from 1 to 100 \msun). 
% b ~ (2.16-2.79)e-12
For the best fit ages of $\sim$ 400--570 Myr the total mass 
of stars formed would then correspond to $\sim (3.6-6.3) \times 10^8$ \msun.
The mass estimated from best fit burst models (of ages $\sim$ 6--20 Myr) is
slightly smaller, $M_\star \sim (0.3 - 1) \times 10^8$ \msun.
% b ~ 8.5e-5 to 2.5e-4
%
If we assume a Salpeter
IMF with \mlow\ $=0.1$ \msun\ the mass and SFR estimates would be higher 
by a factor 2.55, and in good agreement with the values derived by KESR 
and Egami \etal\ (2005).
In all the above cases the total luminosity (unlensed) is typically $L_{\rm bol}
\sim 2 \times 10^{10}$ \lsun.
%%%VERIFIED 2 dec, 2004 - DS

% % % % % % % % % % % % % % % % % % % % % % % % % % % % 

\begin{table*}
%[htb]
\caption{Summary of the main adopted and estimated properties of the analysed high 
redshift galaxies.
The adopted magnification $\mu$ is given in col.\ 2, 
col.\ 3 gives the redshift,
col.\ 4 an indication on the most plausible star formation history (burst or constant SF), 
col.\ 5 a plausible age of the stellar population,
col.\ 6 an estimate of the optical extinction  $A_V$ (for the Calzetti et al.\ 2000 law), 
col.\ 7 the estimated SFR (for a Salpeter IMF from 1--100 \msun), 
col.\ 8 an estimated stellar mass (for same IMF),
col.\ 9 the estimated bolometric luminosity, and
col.\ 10 the estimated \lya\ transmission (fraction of intrinsic emitted flux over observed
\lya\ flux).
See Figs.\ \ref{fig_contour_6a} and \ref{fig_7rev_contour} and the corresponding
text for an estimate of the confidence levels and range of parameters.}
\begin{tabular}{llllllllll}
\hline

Object & $\mu$ & redshift & SF history & age & $A_V$ &  SFR & stellar mass & $L_{\rm bol}$ & \lya\\\ 
       &       &          &            & [Myr] &  [mag]   & [\msun yr$^{-1}$] & [\msun]     & [\lsun] & transmission \\
\hline 
Abell 2218 & 25. & $\sim$ 6.0--7.2 & ? & 3--400 & negligible? & 
			$\sim$ 1 & (0.3--6) $\times$ 10$^8$ & $\sim 2 \times 10^{10}$ \\
KESR \\
\\
Abell 370 & 4.5    & 6.56 & CSFR/ & ? & $\sim$ 1. & 11--41 & (1--4) $\times 10^{8}$ &  (1--4) $\times 10^{11}$ & 23--90 \% \\
HCM6A     &        &      & young burst \\
\\
%\smallskip
         &        &      & composite & young+``old''
%          &        &      & composite population? & young ($\ga$ 10 Myr) + ``old'' ($\ga$ 50-130 Myr)
					& negligible& $>$ 0.4-0.8 & $\sim 2 \times 10^9$& $\sim 3 \times 10^{10}$ 
					& $\ga$ 40 \% \\ 
\hline
%\multicolumn{7}{l}{$^\star$ see the cautionary note on the nature of this object in Sect.\ \ref{s_1916}.}
\end{tabular}
\label{tab_props}
\end{table*}

% % % % % % % % % % % % % % % % % % % % % % % % % % % % % % % % %
%%%%%%%%%%%%%%%%%%%%%%%%%%%%%%%%%%%%%%%%%%%%%%%%%%%%%%%%%%%%%%%%
\section{Results for Abell 370 HCM 6A}
\label{s_370}

\begin{figure*}%[htb]
\centerline{\psfig{figure=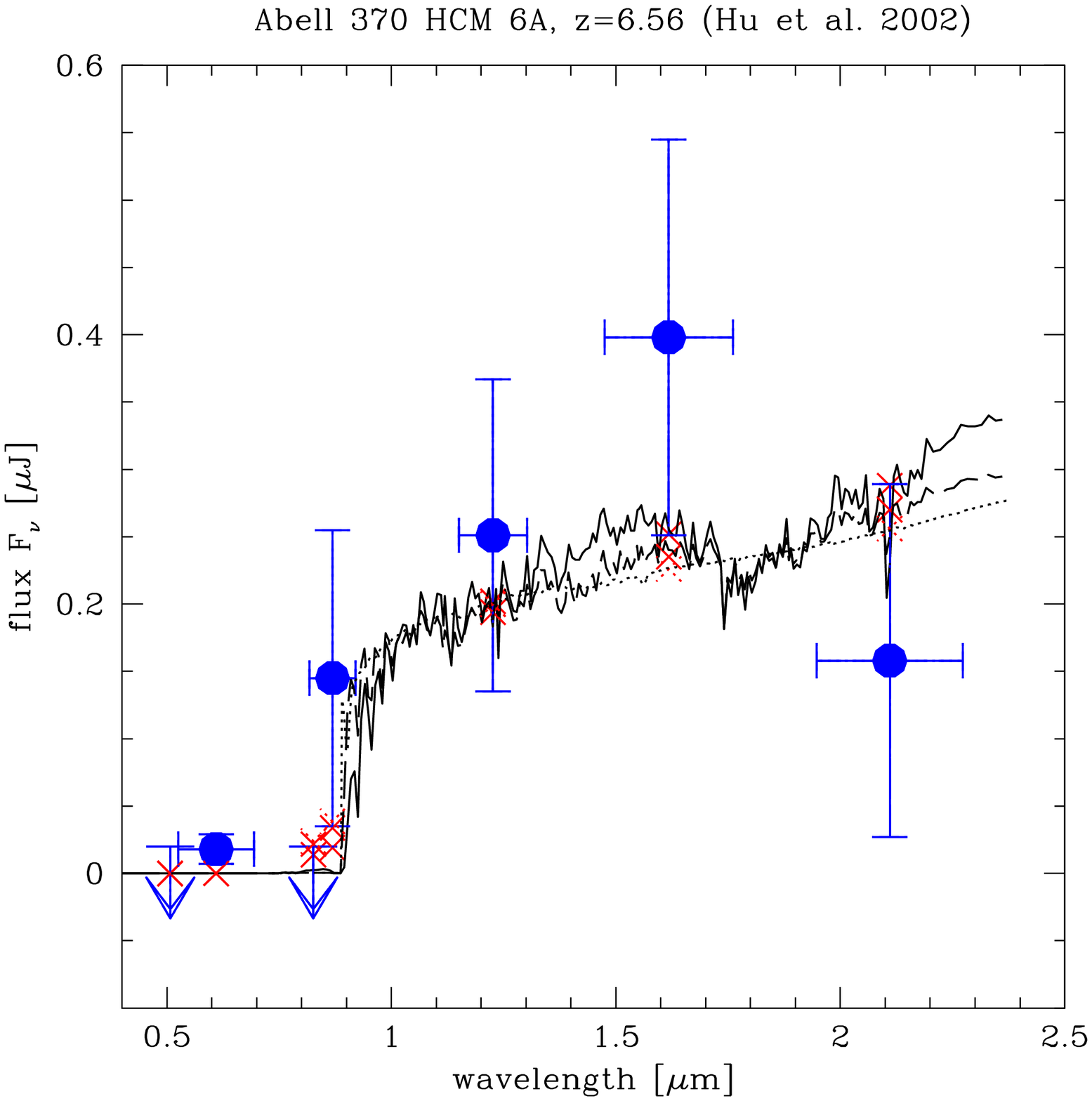,width=8.8cm}
	    \psfig{figure=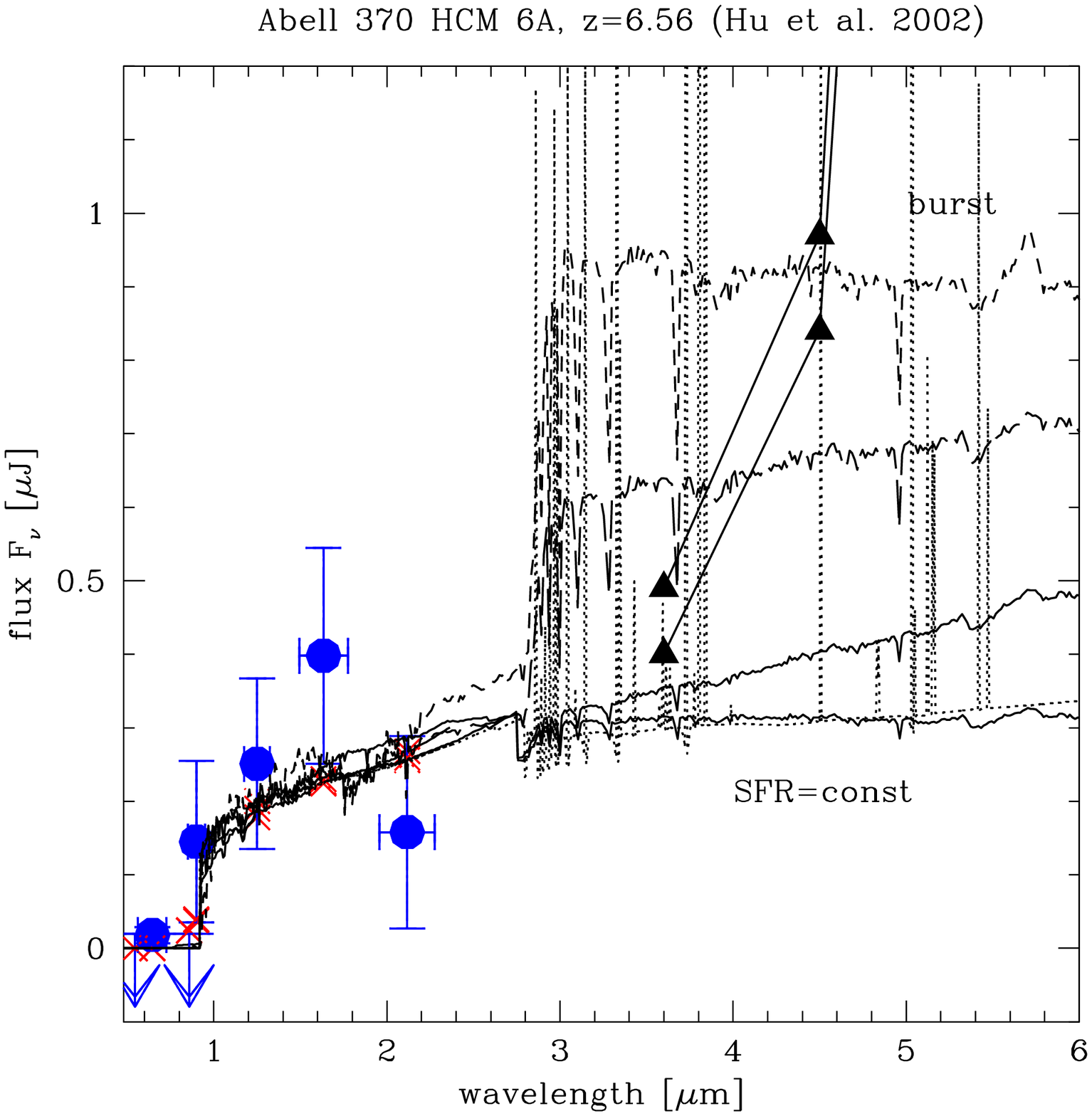,width=8.8cm}}
\caption{Best fits SEDs to the observations of Abell 370 HCM 6A.
The red crosses indicate the corresponding model broad band fluxes.
Solid lines show the best fit for a template from the BC+CWW group,
dotted from SB+QSO group, and dashed from the S03+ group (see explanations
in Sect.\ \ref{s_models}). 
{\bf Left:} Observed spectral range.
{\bf Right:} Predicted SED in Spitzer/IRAC domain for best fit models.
Dashed lines show the bursts from the BCCWW+ and S03+ template groups.
The dotted line is the spectrum of SBS 0335-052 from the SB+QSO group
with additional $A_V=1.$ The solid lines show best fits for constant
star formation using different extinction/attenuation laws (Calzetti 
starburst law versus SMC law). The solid triangles illustrate
the IRAC point-source sensitivity (1 $\sigma$) for low and medium
backgrounds excluding ``confusion noise''.}
\label{fig_sed_6a}
\end{figure*}

\begin{figure*}%[htb]
\centerline{\psfig{figure=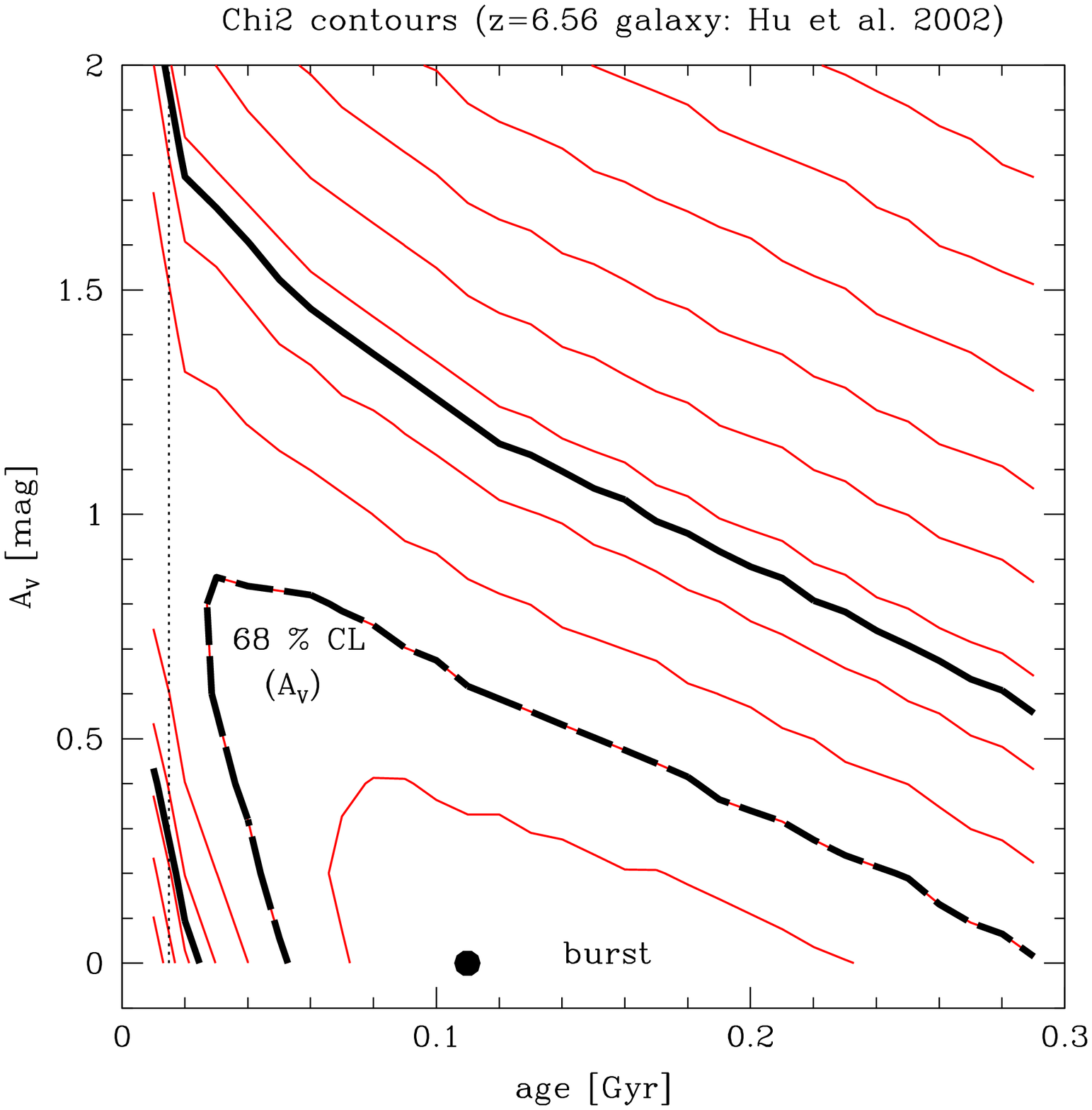,width=8.8cm}
	    \psfig{figure=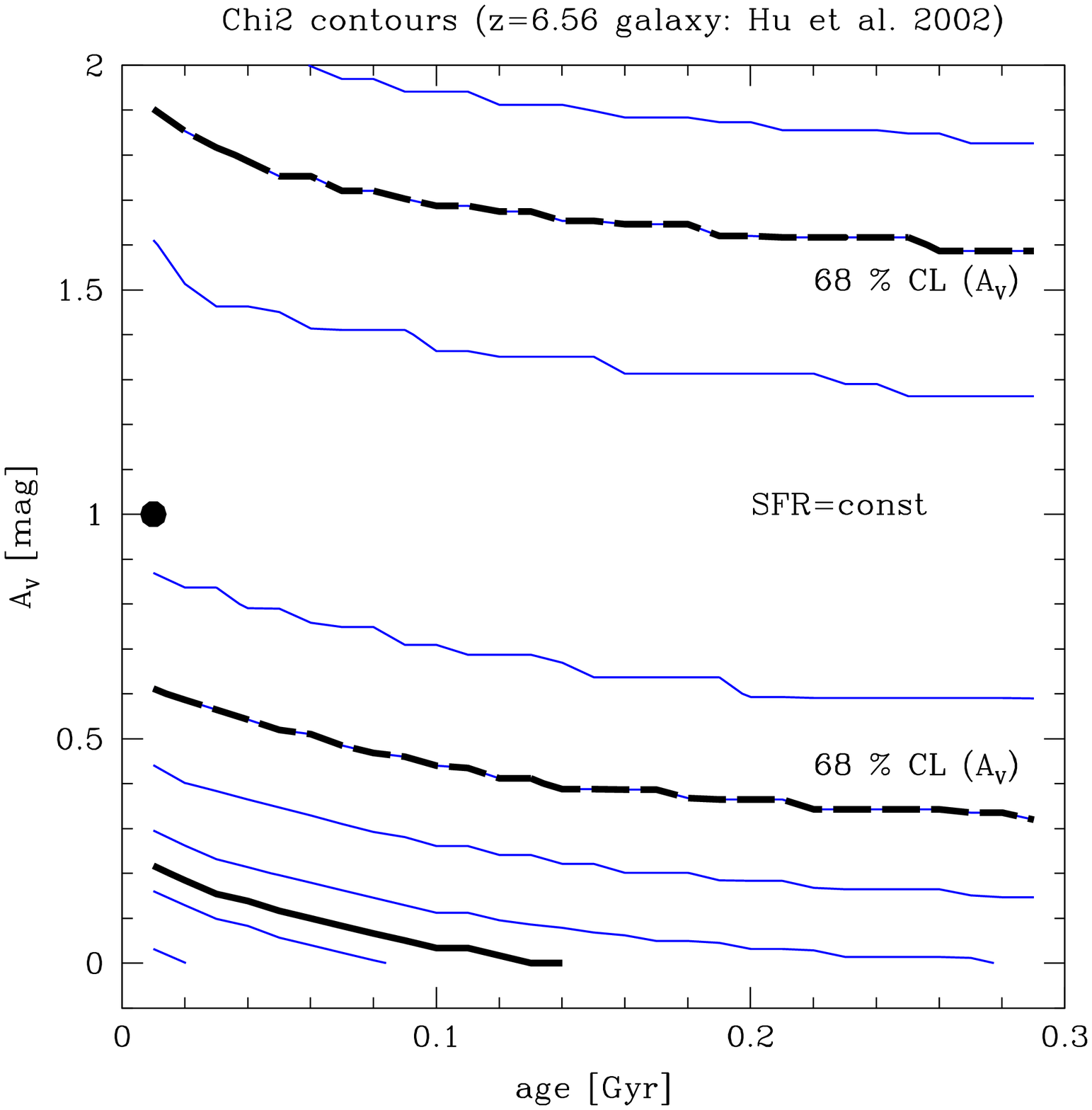,width=8.8cm}}
\caption{$\chi^2$ contour plots showing solutions in extinction -- age 
diagrams. The best solutions are indicated by the black dot.
Equidistant $\chi^2$ levels with a spacing of 0.5 are shown. The 2D 68\% 
confidence region (corresponding to $\Delta \chi^2 = 2.3$)
is delimited by the solid thick black line. The (1D) 68	\% confidence region
for $A_V$ ($\Delta \chi^2 = 1.$) at each given age is delimited by the  dashed thick black line.
{\bf Left:} Plot for solutions using a solar metallicity
burst template from the S03+ template group and the Calzetti attenuation law.
Although providing a good fit to the photometry, the 
region corresponding to ages $\protect\ga$ 15 Myr, right of the dotted vertical line, is excluded
as no emission line would be expected in this case.
{\bf Right:} Same as left panel for constant star formation models.
The solutions indicate a non-negligible extinction, but no constraint on age.
Discussion in text.
}
\label{fig_contour_6a}
\end{figure*}

\subsection{SED fits and inferences on the stellar population}

Overall the published SED of HCM 6A (see Fig.\ \ref{fig_sed_6a})
is ``reddish'', showing an increase of the flux from $Z$ to $H$
and even to $K^\prime$\footnote{The significance of a change of the SED slope
between $ZJH$ and $HK^\prime$ seems weak, and difficult to understand.}.
From this simple fact and the above explanations it is already clear 
qualitatively that one is driven towards solutions with a) ``advanced'' age
and little extinction or b) constant or young star formation 
plus extinction.
However, a) can be excluded as no \lya\ emission would be expected in this case.

Quantitatively, the best solutions obtained for each spectral template group
is shown in the left panel of Figure \ref{fig_sed_6a}.
Indeed, the solutions shown correspond to bursts of ages $\sim$ 50--130 Myr
(BCCWW+, S03 templates) and little or no extinction.
However, as just mentioned, solutions lacking young ($\la$ 10 Myr) massive stars can
be excluded since \lya\ emission is observed.
The best fit empirical SB+QSO template shown corresponds to the spectrum of 
the \hii\ galaxy SBS 0335-052 with an additional extinction of $A_V=1.$
To reconcile the observed SED with \lya, a young population,
e.g.\ such as SBS 0335-052, or constant SF is required. 
In any of these cases fitting the ``reddish'' SED requires
a non negligible amount of reddening.

To illustrate the typical range of possible results we show in
Fig.\ \ref{fig_contour_6a} $\chi^2$ contour maps and corresponding confidence 
intervals for solar metallicity models
(S03+ template group) and reddened with the Calzetti law.
The left panel (burst models) illustrates in particular the need
for progressively higher extinctions the younger the bursts.
%The shaded region is excluded for the absence of \lya\ emission.
All ages $\ga$ 10 Myr are excluded for the absence of \lya\ emission.
From the constant SF models (right panel) we see that for a given
age $A_V$ is typically $\sim$ 0.5--1.8 mag at the 68 \% confidence
level. For obvious reasons, no constraint can be set on the age 
since the onset of (constant) SF.

Hence, from the photometry of HCM 6A and from the presence of \lya\ 
we are led to conclude that this object must suffer from 
reddening with typical values of $A_V \sim 1.$ for a Calzetti
attenuation law.
A somewhat smaller extinction ($A_V \sim 0.4$) can be obtained if 
the steeper SMC extinction law of Pr\'evot et al.\ (1984) 
is adopted. From the present data it is not possible to distinguish
the different extinction/attenuation laws.
Also, it is not possible to draw any constraints on the metallicity of 
HCM 6A from the available data (cf.\ Sect.\ \ref{s_uv}).

What if we are dealing with composite stellar populations?
Indeed it is conceivable that the \lya\ emission originates from a
population of young stars and the ``reddish'' restframe UV flux be due
to another, older population. 
Assuming constant SF, no loss of \lya\ and standard
SFR conversion factors, the observed \lya\ emission implies a maximum
UV flux of the order of 0.1 $\mu$Jy (and approximately constant in $F_\nu$ over
the observed wavelength range: $\lambda_{\rm rest} \ga 3000$ \AA)  
for an unreddened population.
The bulk of the observed flux could then be from an older population.
In this case the rising spectrum from the \zacs\ over the JH (and presumably
K) bands could even be due to an unreddened population;
a strongly increasing flux and probably a significant ``Balmer'' break 
is then expected, similar to the aged burst shown in the right panel of 
Fig.\  \ref{fig_sed_6a}.
This explanation should in principle be testable with Spitzer observations
as discussed below.

How does our possible indication for a high extinction fit in with
other studies?
At redshift $z \la 4$ the extinction of Lyman break galaxies (LBGs)
has been estimated by various authors (e.g.\ Sawicky \& Yee 1998,
Meurer et al.\ 1999,
Adelberger \& Steidel 2000, Shapley \etal\ 2001, Ouchi \etal\ 2004).
Given their mean/median values (typically $<E(B-V)> \sim$ 0.15--0.2) 
and the $E(B-V)$, our finding of a ``high'' extinction is not exceptional. 
Furthermore, Armus \etal\ (1998) find indications for $A_V > 0.5$ mag
from their analysis of a $z=5.34$ galaxy.
However, this implies that 
dust extinction is likely present in starburst galaxies with redshifts above 6.
Large amounts of dust have already been observed in QSOs up to similar
redshift (Bertoldi \etal\ 2003, Walter \etal\ 2003).

\subsection{Properties of HCM6A: SFR, mass, \lya\ transmission}
To estimate properties such as the mass, SFR, and the intrinsic
\lya\ emission from HCM 6A we simply examine the predictions from the
best fit models,
scale them appropriately to the luminosity 
distance\footnote{For the adopted cosmology and $z=6.56$ one
has $d_L=$ 64005.7 Mpc.},
%% 4*pi*d_L^2 = 4.90e+59 cm^2
and correct for the gravitational magnification (here $\mu=4.5$).
The derived quantities are summarised 
in Table \ref{tab_props}.

First we consider single (non-composite) stellar populations.
From the best fit constant SF models (with variable ages) 
we deduce an extinction corrected star formation rate of
the order of SFR(UV) $\sim$ 11 -- 41 \msunyr\ for a Salpeter
IMF from 1 to 100 \msun. For a commonly adopted, although unjustified, 
Salpeter IMF down to 0.1 \msun\ this would increase by a factor 2.55.
Actually this estimate is not very different than the one obtained
from standard SFR calibrations provided the same assumptions on the 
IMF. Indeed the observed restframe UV luminosity,
e.g.\ derived from the average J, H, and K$^\prime$ flux of 
$F_{\rm UV} =(2.6 \pm 0.7) \times 10^{-30}$ \ergscm\ Hz$^{-1}$ 
% typo foiund dec 10$F_{\rm UV} =(2.6 \pm 0.7) \times 10^{-30}$ \ergscm\ Hz$^{-1}$ 
is $L_{\rm rest UV} = 4 \pi d_L^2 F_{\rm UV} / (1+z) / \mu  \approx 4.\times 10^{28}$
\ergs Hz$^{-1}$ (Hu et al.\ 2002), translates to
$SFR_{\rm UV} \approx = c \, L_{\rm rest UV} \, 10^{0.4 A_{\rm UV}}$, where
$c$ is the usual SFR conversion coefficient, and  $A_{\rm UV}$ the UV extinction.
For the standard value $c=1.4 \times 10^{-28}$ from Kennicutt (1998), assuming
a Salpeter IMF down to 0.1 \msun, and 
$A_{\rm UV} \sim$ 2.--3.,
%(cf.\ Sect.\ \ref{s_discuss}),
one has
$SFR \sim$ 35--88 \msunyr; for the IMF used in this work (Salpeter from 1-100 \msun)
this becomes $SFR \sim$ 14--34 \msunyr.
This assumption, and the absence of extinction correction also explains
the difference with $SFR$ estimate of Hu et al.\ (2002)
\footnote{Actually Hu et al.\ (2002) derive without further explanation
$SFR = 9$ \msunyr\ from $L_{\rm rest UV} = 4.\times 10^{28}$ \ergs, whereas the
classical Kennicutt (1998) calibration would yield $SFR = 5.6$ \msunyr\
without extinction correction.}.

For continuous SF over timescales $t_{\rm SF}$ longer than $\sim$ 10 Myr, the total
(bolometric) luminosity output is typically $\sim 10^{10}$ \lsun\ per unit
SFR (in \msunyr) for a Salpeter IMF from 1-100 \msun, quite independently of metallicity. 
The total luminosity  associated with the observed SF is therefore 
$L \sim (1-4) \times 10^{11} \lsun$, close to or just above the limit to 
possibly qualify as a luminous infrared galaxy ($L_{\rm IR} > 10^{11} \lsun$;
cf.\ Sanders \& Mirabel 1996) 
if a significant fraction of its bolometric flux emerges in the (restframe) IR.
%{\b for ***Better LUV/Lbol:
%one has typically intrinsic $Lbol/\lsun \approx 10^{-18} L_\nu$ for the range 1500--2000 \AA,
%quite independently of metallicity and for Mup 30--100 \msun\ (SB99+) ***
%Hence $Lbol \sim (2--6) \times 10^{11} \lsun$ taking a typical extinction
%of 6.3--15.8 in this wavelength range.}
% 
For $t_{\rm SF} \sim$ 10 Myr the estimated stellar mass is 
$M_\star \approx t_{\rm SF} \times SFR \sim (1-4) \times 10^8$ \msun. 

From the data given by Hu et al.\ (2002), the observed \lya\ flux is
$F(\lya) = \mu \times 3.8 \times 10^{-18}$ \ergscm, with the magnification
factor $\mu$.
The \lya\ luminosity per unit SFR from the same S03 models used above is 
$L(\lya)=(2.4-4.4) \times 10^{42}$ \ergs\ (\msunyr)$^{-1}$ for metallicities 
between solar and 1/50 \zsun. The SFR deduced from \lya\ would then be 
SFR$(\lya) \sim$ 0.4--0.8 \msunyr\ for HCM 6A. 
Taking an extinction of $A_V=1.$ (for the Calzetti law) into account
implies a reddening corrected SFR$(\lya) \sim$ 7--12 \msunyr.
The ratio between SFR$(\lya)$/SFR(UV) presumably reflects the incomplete
\lya\ transmission $t_{\rm \lya}$, which can be estimated in various ways.
The most consistent estimate is obtained from the comparison of the predicted
\lya\ luminosity of each best fit model (obtained from fitting the broad-band
SED) to the observed \lya\ luminosity.
From the best fit models with $A_V \sim 1$ and the Calzetti law we obtain
$t_{\rm \lya} \sim$ 23--54 \%; 
in the case of the fit with the Pr\'evot et al.\ extinction law we find
a higher transmission $t_{\rm \lya} \sim$ 90 \%.
For comparison Haiman (2002) assumed a \lya\ transmission of 20 \% from 
the data of Hu et al.
Per definition this \lya\ ``transmission'' corresponds to the ratio 
of the expected/intrinsic \lya\ emission from the starburst over the observed
one. The physical causes for a partial (i.e.\ $<$ 100 \%) transmission are 
of course open to various interpretations (e.g.\ physical processes
destroying \lya\ photons in the host galaxy, absorption in the intervening 
IGM etc.). 
In fact the relatively high \lya\ transmission estimated here could be
somewhat surprising, given the Gunn-Peterson trough observations
in $z \ga 6$ quasars (cf.\ Becker \etal\ 2001, Fan \etal\ 2003)
and the possible presence of dust (this work).

Consider now the case of composite stellar populations.
In this case we retain as a rough estimate in Table \ref{tab_props} the 
$SFR(\lya)$ (from the young population) as a lower limit, 
and the mass and total luminosity is 
derived from the best fit burst model of age $\sim$ 100 Myr assuming
that this ``older'' population dominates the observed continuum flux.
Formally we then have no handle on the \lya\ transmission,
except that it cannot be very low (say $\la$ 40 \% $\approx 0.1 / 0.26
= F_{\rm young}/<F_{\rm obs}>$)
%the ratio of the  $F_{\rm young}/F_{\rm average obs}$) 
since otherwise the associated UV flux from the young population $F_{\rm young}$
would dominate the observed continuum flux $<F_{\rm obs}>$.

\subsection{Spitzer Observatory predictions}
It is interesting to examine the SEDs predicted by the various
models at longer wavelengths, including the rest-frame optical
domain, which is potentially observable with the sensitive IRAC camera
onboard the Spitzer Observatory and other future missions.
In the right panel of Fig.\ \ref{fig_sed_6a} we plot 
the 3 best fits to the observed data for the BCCWW+ and S03+
template groups (``burst'' solutions with no extinction) and the 
SBS 03352-052 template (with additional $A_V=1$) showing 
strong optical emission lines.
We see that these solutions have fluxes comparable to or above 
the detection limit of IRAC/Spitzer
\footnote{The IRAC detection limits plotted here
correspond to the values given by the Spitzer Science Center on 
{\tt http://ssc.spitzer.caltech.edu/irac/sens.html} as 1 $\sigma$ point-source sensitivity 
for low and medium backgrounds for frame times of 200s and described by Fazio \etal\ (2004).
These values do not include ``confusion noise''.}.
On the other hand the strongly reddened constant SF or young burst solutions
do not exhibit a Balmer break and are hence expected to show fluxes
just below the IRAC sensitivity at 3.6 \micron\ and significantly
lower at longer wavelengths.
As \lya\ emission is expected only for the reddened SEDs the
latter solutions (low 3.6-4.5 \micron\ flux) are predicted to apply to HCM 6A,
except if composite stellar populations are considered.
Indeed, a high  3.6 and 4.5 \micron\ flux could be a good indication 
for a composite stellar population, as discussed above.
In addition it is important to secure higher accuracy photometry
especially in the near-IR (JHK) to assess the accuracy of the redward increasing 
shape of the spectrum (in $F_\nu$), which drives one towards solutions 
with non-negligible extinction.

%%%%%%%%%%%%%%%%%%%%%%%%%%%%%%%%%%%%%%%%%%%%%%%%%%%%%%%%%%%%%%%%
\section{Conclusion}
\label{s_conclude}
Using SED fitting techniques considering a large
number of parameters (mainly a vast library of empirical and theoretical 
template spectra, variable extinction and extinction laws) 
we have attempted to constrain the properties of the stellar populations
and \lya\ emission of two strongly lensed galaxies with redshifts $z \ga$ 6
from their observed SED including various ground-based observations,
HST, and Spitzer observations.
 
The following main results 
have been obtained for these objects
(see Sects.\ \ref{s_370}, \ref{s_kesr}),
% and \ref{s_discuss}, 
and summary in Table \ref{tab_props}): 

\begin{itemize} 
\item {\bf Triple arc in Abell 2218} discovered by Kneib \etal\ (2004, KESR).
The most likely redshift of this source is $z \sim$ 6.0--7.2 
taking into account both our photometric determination and lensing 
considerations.

SED fits indicate generally a low extinction ($E(B-V) \la 0.05$)
but do not strongly constrain the SF history. 
Best fits have typical ages of $\sim$ 3 to 400 Myr. A reasonable
maximum age of (250--650) Myr (1 $\sigma$ interval) can be estimated. 
However, the apparent 4000 \AA\ break 
observed recently from combination of IRAC/Spitzer and HST observations,
can also equally well be reproduced with the template of a young
($\sim$ 3--5 Myr) burst where strong restframe optical emission lines 
enhance the 3.6 and 4.5 \micron\ fluxes.

The estimated SFR is typically $\sim$ 1 \msunyr\ for a Salpeter IMF
from 1-100 \msun, in agreement with previous estimates.

Given the poor constraint on age and SF history,
we conclude that intrinsic \lya\ emission
may or may not be present in this galaxy. 
The apparent non-detection of \lya\ by KESR can therefore 
even be understood without invoking \lya\ destruction.

\item {\bf Abell 370 HCM 6A} discovered by Hu \etal\ (2002).
The relatively red SED and the presence of \lya\ emission indicate
basically two possible solutions:
1) a young burst or ongoing constant SF with non-negligible extinction
($A_V \sim$ 0.5--1.8 at a 1 $\sigma$ level) 
or 2) a composite young + ``old''  stellar population.

For the first case,
best fits are obtained for constant SF with $E(B-V) \sim 0.25$.
In consequence previous SFR estimates for this source must likely be revised
upward. 
%To the best of our knowledge this represents the first possible evidence
%for dust extinction in a galaxy with $z>6$.
%Taking into account the likely presence of dust, 
If correct, the bolometric luminosity 
of this galaxy is estimated to be $L \sim (1-4) \times 10^{11} \lsun$,
comparable to the luminosity of infrared luminous galaxies.
Furthermore a \lya\ transmission of $\sim$ 23--90 \% is estimated from our best fit models.

Alternatively the observed 0.9-2.2 \micron\ SED could also be fit 
without extinction by a composite ``young'' and ``old'' stellar population,
where the former would be responsible for the \lya\ emission and a fraction
of the restframe UV flux. 
The SFR, stellar mass, and total luminosity are then lower than in case 1.
The two scenarios may be distinguishable with IRAC/Spitzer observations at
3.6 and 4.5 \micron.

Given the limited observed spectral range, the present data does not allow 
to draw any firm constraints on the maximum
age of the stellar population. 

\end{itemize} 

In general it should also be noted that broad-band SED fits or measurements
of the UV slope do not allow one to determine the metallicity of 
a star forming galaxy from a theoretical point of view and in terms 
of individual objects given important degeneracies 
(cf.\ Sect.\ \ref{s_uv}).

The estimates of the \lya\ transmissions presented here can 
in principle be used to constrain the intervening IGM properties,
and therefore probe the reionisation of the Universe.
%(see Sect.\ \ref{s_discuss}).

Although the results obtained here from this exploratory study
of just two lensed galaxies, the highest known redshift galaxies
with photometric detections in at least 3--4 filters, cannot provide 
a general view on the SF and IGM properties at $z \ga 6$,
there is good hope that the sample of such objects will
considerably increase in the near future with the availability
of large ground-based telescopes and sensitive space borne
observatories such as Spitzer and even more so with the planned
JWST.

%%%%%%%%%%%%%%%%%%%%%%%%%%%%%%%%%%%%%%%%%%%%%%%%%%%%%%%%%%%%%%%%
\section*{Acknowledgements}

We thank an anonymous referee for critical comments which helped
to improve the paper.
We thank Eichi Egami and Johan Richard for comments on the HST and 
Spitzer photometry of the arc in Abell 2218 KESR, Jean-Paul Kneib for
information on Keck spectroscopy of this object, and Yuri Izotov for
communicating spectra of metal-poor galaxies.
Part of this work was supported by the Swiss National Science Foundation 
and the CNRS.

%%%%%%%%%%%%%%%%%%%%%%%%%%%%%%%%%%%%%%%%%%%%%%%%%%%%%%%%%%%%%%%%

\end{document}